\documentclass[epj]{webofc}
\usepackage[varg]{txfonts}   
\wocname{EPJ Web of Conferences}
\woctitle{CONF12}
\begin{document}
\selectlanguage{english}
\title{Holographic QCD for H-dibaryon (uuddss)}
%
%

\author{Hideo Suganuma\inst{1}\fnsep\thanks{\email{suganuma@scphys.kyoto-u.ac.jp}} \and
        Kohei Matsumoto\inst{2} 
}

\institute{
Department of Physics, 
Kyoto University, Kitashirakawaoiwake, Sakyo, Kyoto 606-8502, Japan
\and
Yukawa Institute for Theoretical Physics (YITP), 
Kyoto University, Sakyo, Kyoto 606-8502, Japan
}

\abstract{
The H-dibaryon (uuddss) is studied in holographic QCD for the first time. 
In holographic QCD, 
four-dimensional QCD, i.e., SU($N_c$) gauge theory with chiral quarks, 
can be formulated with $S^1$-compactified D4/D8/$\overline{\rm D8}$-brane system. 
In holographic QCD with large $N_c$, all the baryons appear as topological chiral solitons 
of Nambu-Goldstone bosons and (axial) vector mesons, and the H-dibaryon can be described as an SO(3)-type topological soliton with $B=2$.
We derive the low-energy effective theory to describe the H-dibaryon in holographic QCD.
The H-dibaryon mass is found to be twice of the $B=1$ hedgehog-baryon mass,
$M_{\rm H} \simeq 2.00 M_{B=1}^{\rm HH}$,
and is estimated about 1.7GeV, 
which is smaller than mass of two nucleons (flavor-octet baryons), in the chiral limit. 
}
\maketitle

\section{Introduction}

Nowadays, QCD is established as the fundamental theory of the strong interaction, 
and all the experimentally observable hadrons have been considered as 
color-singlet composite particles of quarks and gluons. 
From QCD, as well as ordinary mesons ($\bar{q}q$) and baryons ($qqq$) 
in the valence picture, there can exist ``exotic hadrons'' \cite{IP85} 
such as glueballs, multi-quarks \cite{J77,OST05} and hybrid hadrons,   
and the exotic-hadron physics has been an interesting field theoretically 
and experimentally. 

The H-dibaryon, $B=2$ SU(3) flavor-singlet bound state of uuddss, has been 
one of the oldest multi-quark candidates, first predicted by R.~L.~Jaffe in 1977 
from a group-theoretical argument of the color-magnetic interaction 
in the MIT bag model \cite{J77}.
In 1985, the H-dibaryon was also investigated \cite{BLR85,JK85} 
in the Skyrme-Witten model \cite{S61,W79,ANW83}. 
These two model calculations suggested a low-lying H-dibaryon below the 
$\Lambda\Lambda$ threshold, which means the stability of H against the strong decay.
In 1991, however, Imai group experimentally excluded the low-lying H-dibaryon \cite{I91},  
and found the first event of the double hyper nuclei, 
i.e., $_{\Lambda \Lambda}^{~~6}{\rm He}$, instead. 
Then, the current interest is mainly possible existence of the H-dibaryon 
as a resonance state. 

Theoretically, it is still interesting to consider the stability of H-dibaryons 
in the SU(3) flavor-symmetric case of $m_u=m_d=m_s$ \cite{NPL11,HAL11,YH16}, 
because the large mass of H may be due to an SU(3) flavor-symmetry breaking 
by the large s-quark mass, $m_s \gg m_{u,d}$, in the real world. 
Actually, recent lattice QCD simulations suggest the stable H-dibaryon 
in an SU(3) flavor-symmetric and large quark-mass region \cite{NPL11,HAL11}. 

So, how about the H-dibaryon in the chiral limit of $m_u=m_d=m_s=0$? 
Although the lattice QCD calculation is usually a powerful method to evaluate hadron masses, 
it is fairly difficult to take the chiral limit, because a large-volume lattice is needed for 
such a calculation to control massless pions. 

In this paper, we study the H-dibaryon and its properties in the chiral limit 
using holographic QCD \cite{MNS16}, which has a direct connection to QCD, 
unlike most effective models.
In particular, we investigate the H-dibaryon mass from the viewpoint of its stability 
in the chiral limit.

\section{Holographic QCD}

In this section, we briefly summarize the construction of holographic QCD from a D-brane system \cite{W98,SS05}, and derive the low-energy effective theory of QCD \cite{NSK07}   
at the leading order of $1/N_c$ and 1/$\lambda$ expansions, 
where the 't~Hooft coupling $\lambda \equiv N_c g_{\rm YM}^2$ is given 
with the gauge coupling $g_{\rm YM}$.

\subsection{QCD-equivalent D-brane system}

\begin{figure}[h]
\centering
\sidecaption
\includegraphics[width=80mm,clip]{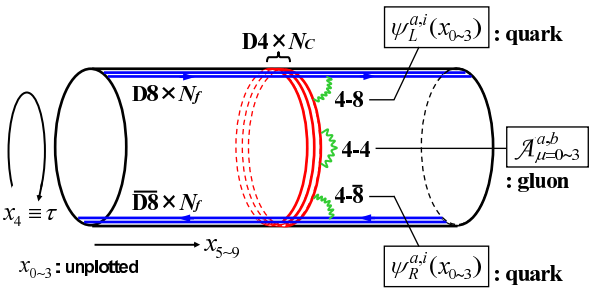}
\caption{Construction of holographic QCD with 
an $S^1$-compactified D4/D8/$\overline{\rm D8}$-brane system,  
which corresponds to non-SUSY four-dimensional QCD with chiral quarks 
\cite{SS05,NSK07}. This figure is taken from Ref.\cite{NSK07}.
}
\label{fig-1}
\end{figure}

Just after J. M. Maldacena's discovery of the AdS/CFT correspondence in 1997 \cite{M97}, 
E.~Witten \cite{W98} succeeded in 1998 the formulation of 
non-SUSY four-dimensional pure SU($N_c$) gauge theories 
using an $S^1$-compactified D4-brane in the superstring theory. 
In 2005, Sakai and Sugimoto showed a remarkable formulation of  
four-dimensional QCD, i.e., SU($N_c$) gauge theory with chiral quarks, 
using an $S^1$-compactified D4/D8/$\overline{\rm D8}$-brane system \cite{SS05}, 
as shown in Fig.~\ref{fig-1}.
Such a construction of QCD is often called holographic QCD.

This QCD-equivalent D-brane system consists of $N_c$ D4-branes and $N_f$ D8/$\overline{\rm D8}$-branes, which give color and flavor degrees of freedom, respectively.
In this system, gluons appear as 4-4 string modes on $N_c$ D4-branes, 
and the left/right quarks appear as 4-8/4-$\bar 8$ string modes 
at the cross point between D4 and D8/$\overline{\rm D8}$ branes, 
as shown in Fig.~\ref{fig-1}.
This D-brane system possesses the ${\rm SU}(N_c)$ gauge symmetry and 
the exact chiral symmetry \cite{SS05}, and gives QCD in the chiral limit. 

\begin{figure}[h]
\centering
\sidecaption
\includegraphics[width=80mm,clip]{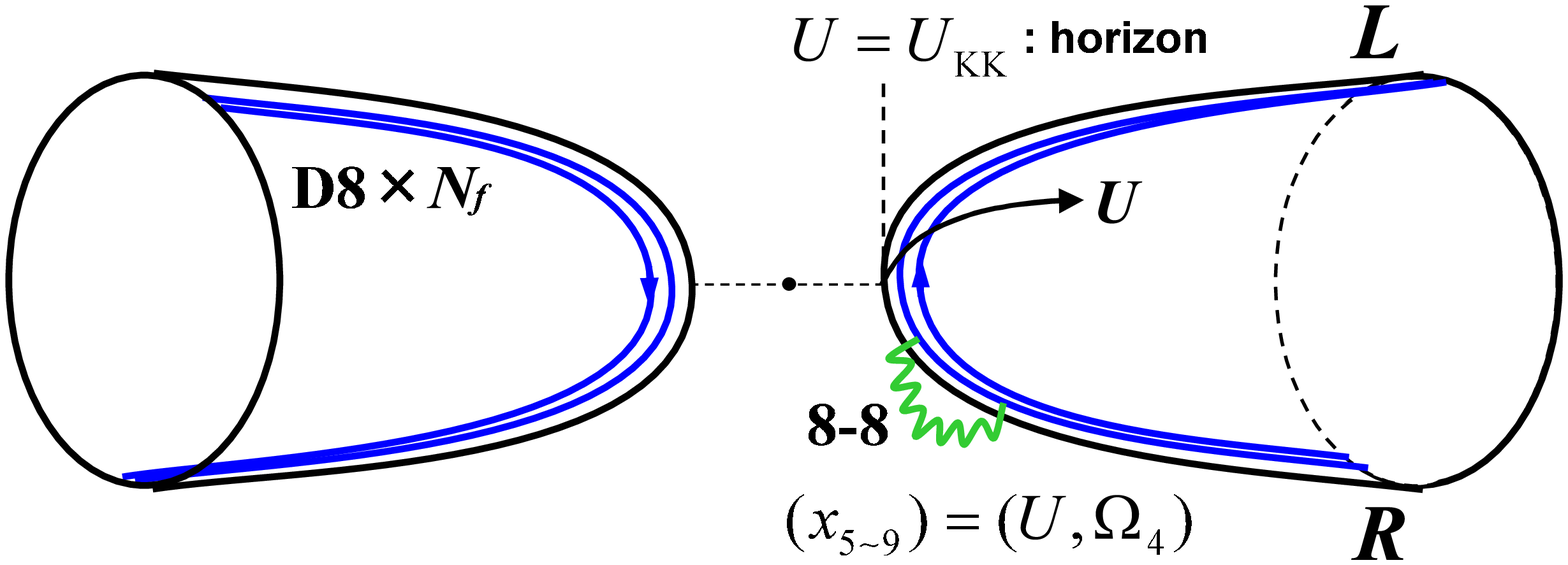}
\caption{Holographic QCD after the replacement of 
large-$N_c$ D4 branes by a gravitational background 
via the gauge/gravity correspondence \cite{W98,SS05,NSK07}. \newline
This figure is taken from Ref.\cite{NSK07}.
}
\label{fig-2}
\end{figure}

In holographic QCD, $1/N_c$ and $1/\lambda$ expansions are usually taken.
In large $N_c$, D4-branes are the dominant gravitational source, and can be 
replaced by their SUGRA solution \cite{SS05}  
as shown in Fig.~\ref{fig-2}, via the gauge/gravity correspondence.
In large $\lambda$, the strong-coupling gauge theory is converted into 
a weak-coupling gravitational theory \cite{W98}.
In this paper, we consider the leading order of $1/N_c$ and $1/\lambda$ expansions.

\subsection{Low-energy effective theory}

In the presence of the D4-brane gravitational background $g_{MN}$, 
the D8/$\overline{\rm D8}$ brane system can be expressed 
with the non-Abelian Dirac-Born-Infeld (DBI) action, 
\begin{equation}
S^{\rm DBI}_{\rm D8} = T_8 \int d^9x \:e^{-\phi} \sqrt{-{\rm det}(g_{MN}+2\pi \alpha' F_{MN})} \:,
\end{equation}
at the leading order of $1/N_c$ and $1/\lambda$ expansions. 
Here, $F_{MN} \equiv \partial_M A_N-\partial_N A_M +i[A_M, A_N]$ 
is the field strength of the U($N_f$) gauge field $A_M$ 
in the flavor space on the D8 brane.
The surface tension $T_8$, the dilaton field $\phi$ and 
the Regge slope parameter $\alpha'$ 
are defined in the framework of the superstring theory, and, 
for the simple notation, we have taken the $M_{\rm KK} = 1$ unit, where 
the Kaluza-Klein mass $M_{\rm KK}$ is the energy scale of this theory \cite{SS05}.

After some calculations, 
one can derive the meson theory equivalent to infrared QCD 
at the leading order of $1/N_c$ and $1/\lambda$ \cite{SS05,NSK07}. 
For the construction of the low-energy effective theory, 
we only consider massless Nambu-Goldstone (NG) bosons 
and the lightest SU($N_f$) vector meson 
$\rho_\mu(x) \equiv \rho_\mu(x)^a T^a \in {\rm su}(N_f)$, 
which we simply call ``$\rho$-meson".
We eventually derive the four-dimensional effective action 
in Euclidean space-time $x^\mu=(t, {\bf x})$ \cite{NSK07},
\begin{align}
S_{\rm HQCD} = \int d^4x \:&\Bigl\{
\: \frac{f_{\pi}^2}{4} \mathrm{tr}(L_{\mu}L_{\mu}) 
- \frac{1}{32e^2} \mathrm{tr} [L_{\mu}, L_{\nu}]^2 
+ \frac{1}{2} \mathrm{tr}(\partial_{\mu}\rho_{\nu} - \partial_{\nu}\rho_{\mu} )^2
+ m_{\rho}^2 \mathrm{tr}(\rho_{\mu}\rho_{\mu}) \notag \\
&- i g_{3\rho}\mathrm{tr} \bigl\{ (\partial_{\mu}\rho_{\nu} - \partial_{\nu}\rho_{\mu} )[\rho_{\mu}, \rho_{\nu}] \bigr\} 
- \frac{1}{2}g_{4\rho} \mathrm{tr} [\rho_{\mu}, \rho_{\nu}]^2 
+i g_1 \mathrm{tr} \bigl\{ [\alpha_{\mu}, \alpha_{\nu}] (\partial_{\mu}\rho_{\nu} - \partial_{\nu}\rho_{\mu} ) \bigr\} \notag \\
&+ g_2 \mathrm{tr} \bigl\{ [\alpha_{\mu}, \alpha_{\nu}] [\rho_{\mu}, \rho_{\nu}] \bigr\} 
+ g_3 \mathrm{tr} \bigl\{ [\alpha_{\mu}, \alpha_{\nu}] ([\beta_{\mu}, \rho_{\nu}] +[\rho_{\mu}, \beta_{\nu}] ) \bigr\} \notag \\
&-i g_4 \mathrm{tr} \bigl\{ (\partial_{\mu}\rho_{\nu} - \partial_{\nu}\rho_{\mu} ) ([\beta_{\mu}, \rho_{\nu}] + [\rho_{\mu}, \beta_{\nu}] ) \bigr\} 
- g_5 \mathrm{tr} \bigl\{ [\rho_{\mu}, \rho_{\nu}] ([\beta_{\mu}, \rho_{\nu}] +[\rho_{\mu}, \beta_{\nu}] ) \bigr\} \notag \\
&- \frac{1}{2} g_6 \mathrm{tr} \bigl([\alpha_{\mu}, \rho_{\nu}] + [\rho_{\mu}, \alpha_{\nu}] \bigr)^2 
- \frac{1}{2} g_7 \mathrm{tr} \bigl([\beta_{\mu}, \rho_{\nu}] + [\rho_{\mu}, \beta_{\nu}] \bigr)^2
\: \Bigr\},
\label{HQCDaction}
\end{align}
where $L_{\mu}$ is defined with the chiral field $U(x)$ or 
the NG boson field $\pi(x) \equiv \pi^a(x)T^a \in {\rm su}(N_f)$ as
\begin{equation}
L_{\mu} \equiv \frac{1}{i}U^{\dagger}\partial_{\mu}U \in {\rm su}(N_f), \quad
U(x) \equiv e^{i2\pi(x)/f_{\pi}} \in {\rm SU}(N_f).
\end{equation}
The axial vector current $\alpha_{\mu}$ and 
the vector current $\beta_{\mu}$ are defined as 
\begin{equation}
\alpha_{\mu} \equiv l_{\mu} - r_{\mu} \in {\rm su}(N_f)_A, \quad
\beta_{\mu} \equiv \frac{1}{2} (l_{\mu} + r_{\mu}) \in {\rm su}(N_f)_V,
\end{equation}
with the left and the right currents,
\begin{equation}
l_{\mu} \equiv \frac{1}{i}\xi^{\dagger}\partial_{\mu}\xi, \quad 
r_{\mu} \equiv \frac{1}{i}\xi\partial_{\mu}\xi^{\dagger}, \quad
\xi(x) \equiv e^{i\pi(x)/f_{\pi}} \in {\rm SU}(N_f).
\end{equation}

Thus, we obtain the effective meson theory derived from QCD in the chiral limit 
at the leading order of $1/N_c$ and $1/\lambda$ expansions.  
Note that this theory has just two independent parameters, 
e.g., the Kaluza-Klein mass $M_{\rm KK} \sim$ 1GeV 
and $\kappa \equiv \lambda N_c/216\pi^3$ \cite{SS05,HSSY07}, 
and all the coupling constants and masses in the effective action (\ref{HQCDaction}) 
are expressed with them \cite{NSK07}. 
As a remarkable fact, in the absence of the $\rho$-meson, 
this effective theory reduces to the Skyrme-Witten model \cite{S61} 
in Euclidean space-time, 
\begin{equation}
{\cal L}_{\rm Skyrme} = \frac{f_{\pi}^2}{4} {\rm tr}(L_{\mu}L_{\mu}) - \frac{1}{32e^2}
{\rm tr}[L_{\mu},L_{\nu}]^2.
\end{equation}

\section{H-dibaryon as a B=2 Topological Chiral Soliton in Holographic QCD}

As a general argument, large-$N_c$, QCD becomes a weakly interacting meson theory, 
and baryons are described as topological chiral solitons of mesons \cite{W79}.
In holographic QCD with large $N_c$, the H-dibaryon is also described 
as a $B=2$ chiral soliton, 
and its static profile is expressed with the ``SO(3)-type hedgehog Ansatz'', 
similarly in the Skyrme-Witten model \cite{BLR85,JK85}.
Here, the SO(3) is the flavor-symmetric subalgebra of SU(3)$_f$, 
and its generators $\Lambda_{i=1,2,3}$ are 
\begin{equation}
\Lambda_1=\lambda_7=
\begin{pmatrix}
0 & 0 & 0 \\
0 & 0 & -i \\
0 & i & 0 \\
\end{pmatrix},\quad
\Lambda_2=-\lambda_5=
\begin{pmatrix}
0 & 0 & i \\
0 & 0 & 0 \\
-i & 0 & 0 \\
\end{pmatrix},\quad
\Lambda_3=\lambda_2=
\begin{pmatrix}
0 & -i & 0 \\
i & 0 & 0 \\
0 & 0 & 0 \\
\end{pmatrix},
\end{equation}
which satisfy the SO(3) algebra and the following relations,  
\begin{equation}
[\Lambda_i, \Lambda_j] = i\epsilon_{ijk}\Lambda_k, 
\quad
({\bf \Lambda \cdot \hat{x}})^3={\bf \Lambda \cdot \hat{x}},
\quad
{\rm Tr}[({\bf \Lambda \cdot \hat{x}})^2-2/3]=0, 
\label{SO3AL}
\end{equation}
with $\ \hat{\bf x} \equiv {\bf x}/r$ and $r \equiv |{\bf x}|$.
The SO(3)-type hedgehog Ansatz \cite{BLR85,JK85,MNS16} is generally expressed as 
\begin{equation}
U({\bf x}) = e^{ i\{ ({\bf \Lambda \cdot \hat{x}} )F(r) + 
[( {\bf \Lambda \cdot \hat{x}} )^2 - 2/3 ] \varphi(r) \} } 
\in {\rm SU(3)}_f, 
\quad F(r) \in {\bf R}, 
\quad \varphi(r) \in {\bf R},
\label{SO3HH}
\end{equation}
where $F(r)$ and $\varphi(r)$ are the chiral profile functions 
characterizing the NG boson field. 
Note that $U({\bf x})$ in Eq.(\ref{SO3HH}) is the general form of the special unitary matrix 
which consists of ${\bf \Lambda \cdot \hat{x}}$, 
because of Eq.(\ref{SO3AL}).
For the topological soliton, the $B=2$ boundary condition \cite{BLR85,JK85} is given as 
\begin{equation}
F(\infty)=\varphi(\infty)=0, \quad F(0)=\varphi(0)=\pi.
\label{BC}
\end{equation}
%
On the SU(3)$_f$ $\rho$-meson field, 
we use the SO(3) Wu-Yang-'t~Hooft-Polyakov Ansatz, 
\begin{equation}
\rho_0({\bf x})=0,\quad \rho_i({\bf x})
= \epsilon_{ijk} \hat{x_j} G(r) \Lambda_k \in {\rm so}(3) \subset {\rm su}(3),
\quad G(r) \in {\bf R},
\label{SO3WY}
\end{equation}
similarly in the $B=1$ case in holographic QCD \cite{NSK07}. 
(This $G(r)$ corresponds to $-\tilde G(r)$ in Ref.\cite{NSK07}.)
Thus, all the above treatments are symmetric in the (u, d, s) flavor space. 

Substituting Ans\"atze (\ref{SO3HH}) and (\ref{SO3WY}) in Eq.(\ref{HQCDaction}), 
we derive the effective action to describe the static H-dibaryon 
in terms of the profile functions $F(r)$, $\varphi(r)$ and $G(r)$ \cite{MNS16}: 
\begin{align}
S_{\rm HQCD} = \int d^4x &\:\biggl\{
\:\: \frac{f_{\pi}^2}{4} \Bigl[\frac{2}{3}\varphi'^2+2F'^2+\frac{8}{r^2}(1 - \cos F \cos \varphi ) \Bigr] \notag 
+ \frac{1}{32e^2} \frac{16}{r^2} \Bigl[(\varphi'^2 + F'^2)(1 - \cos F \cos \varphi ) \\
&+2\varphi' F' \sin F \sin \varphi 
+ \frac{1}{r^2} \bigl\{(1 - \cos F \cos \varphi )^2 
+ 3\sin^2 F \sin^2 \varphi \bigr\}\Bigr] \notag \\
&+ \frac{1}{2} \Bigl[8 \Bigl( \frac{3}{r^2} G^2 + \frac{2}{r} GG' + G'^2 \Bigr)\Bigr]
+ m_{\rho}^2 [4 G^2] 
+ g_{3\rho}\Bigl[8\frac{G^3}{r}\Bigr] 
+ \frac{1}{2}g_{4\rho}[4G^4] \notag\\ 
&- g_1 \Bigl[ \frac{16}{r} \Bigl\{ \Bigl(\frac{1}{r}G+G' \Bigr) \Bigl(F'\sin \frac{F}{2}\cos \frac{\varphi}{2}+ \varphi' \cos \frac{F}{2} \sin \frac{\varphi}{2} \Bigr) + \frac{1}{r^2}G (1-\cos F \cos \varphi) \Bigr\} \Bigr] \notag\\
&- g_2 \Bigl[\frac{8}{r^2}G^2(1 - \cos F \cos \varphi )\Bigr] \notag\\
&+ g_3 \Bigl[ \frac{16}{r^3} G \Bigl\{ 3\sin F \sin \frac{F}{2} \sin \varphi \sin \frac{\varphi}{2} + \Bigl(1-\cos \frac{F}{2} \cos \frac{\varphi}{2}\Bigr)(1-\cos F \cos \varphi) \Bigr\}\Bigr] \notag \\
&- g_4 \Bigl[ \frac{16}{r^2} G^2 \Bigl(1-\cos \frac{F}{2} \cos \frac{\varphi}{2}\Bigr) \Bigr] 
- g_5 \Bigl[\frac{8}{r}G^3 \Bigl(1-\cos \frac{F}{2} \cos \frac{\varphi}{2}\Bigr) \Bigr] \notag \\
&+ g_6 \bigl[4G^2(F'^2+\varphi'^2)\bigr] 
+g_7 \Bigl[ \frac{8}{r^2} G^2 \Bigl\{ 3\sin^2 \frac{F}{2} \sin^2 \frac{\varphi}{2} + \Bigl(1-\cos \frac{F}{2} \cos \frac{\varphi}{2} \Bigr) ^2 \Bigr\} \Bigr]
\: \biggr\} \notag \\
=\int dt &\int_0^\infty dr~4\pi r^2 \varepsilon[F(r), \varphi(r), G(r)]. 
\label{HQCDFG}
\end{align}

\section{H-dibaryon Solution in Holographic QCD}

To obtain the topological soliton solution of the H-dibaryon in holographic QCD, 
we numerically calculate the profiles $F(r)$, $\varphi(r)$ and $G(r)$ 
\cite{MNS16} by minimizing the Euclidean effective action (\ref{HQCDFG}) 
under the boundary condition (\ref{BC}) \cite{SS98}. 
The two independent parameters, e.g.,   
$M_{\rm KK}$ and $\kappa \equiv \lambda N_c/216\pi^3$, 
are set to reproduce the pion decay constant $f_{\pi}$=92.4MeV and 
the $\rho$-meson mass $m_{\rho}$=776MeV \cite{SS05, NSK07}.

\begin{figure}[h]
\centering
\sidecaption
\includegraphics[width=65mm,clip]{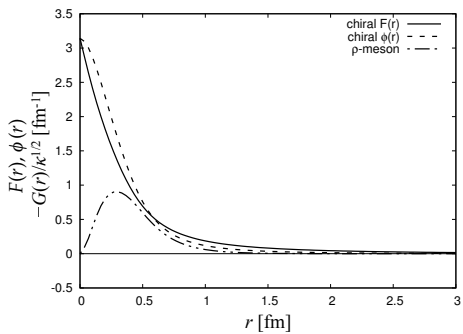}
\caption{The chiral profiles, $F(r)$ and $\varphi(r)$, 
and the scaled $\rho$-meson profile $G(r)/\kappa^{1/2}$ 
in the H-dibaryon as the SO(3)-type hedgehog soliton solution 
in holographic QCD.  Here, the topological boundary condition of $B=2$ is  
$F(0)=\varphi(0)=\pi$ and $F(\infty)=\varphi(\infty)=0$.}
\label{fig-3}
\end{figure}

\begin{figure}[h]
\centering
\sidecaption
\includegraphics[width=65mm,clip]{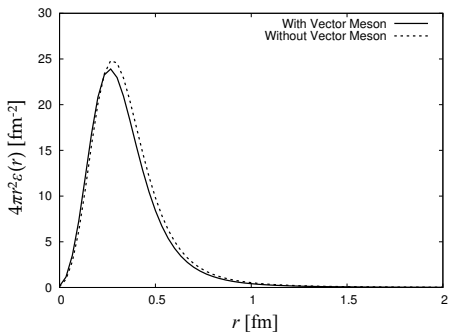}
\caption{The energy density distribution $4\pi r^2 \varepsilon(r)$ 
in the H-dibaryon (solid curve), 
and that without vector mesons (dashed curve) for comparison.}
\label{fig-4}
\end{figure}

For the H-dibaryon solution in holographic QCD, 
we obtain the chiral profiles, $F(r)$ and $\varphi(r)$, and 
the scaled $\rho$-meson profile $G(r)/\kappa^{1/2}$ as shown in Fig.~\ref{fig-3}, 
and estimate the H-dibaryon mass of 
$M_{\rm H} \simeq 1673 {\rm MeV}$ in the chiral limit. 
Figure~\ref{fig-4} shows the energy density $4\pi r^2 \varepsilon(r)$  
in the H-dibaryon.
The root mean square radius of the H-dibaryon is estimated as 
$\sqrt{\langle r^2 \rangle_{\rm H}} \simeq$ 0.413fm 
in terms of the energy density.
For comparison, we calculate the $B=1$ hedgehog (HH) baryon 
in holographic QCD with the same numerical condition, and estimate   
$M_{B=1}^{\rm HH} \simeq 836.7 {\rm MeV}$ and 
$\sqrt{\langle r^2 \rangle_{B=1}^{\rm HH}} \simeq 0.362 {\rm fm}$.
Thus, the H-dibaryon mass is twice of the $B=1$ hedgehog-baryon mass,
$M_{\rm H} \simeq 2.00 M_{B=1}^{\rm HH}$.

We summarize in Table~\ref{tab-1} the mass and the radius of the H-dibaryon 
and the $B=1$ hedgehog baryon in holographic QCD.
Since the nucleon mass $M_{\rm N}$ is larger than 
the $B=1$ hedgehog mass $M_{B=1}^{\rm HH}$ 
by the rotational energy \cite{S61,ANW83}, 
the H-dibaryon mass is smaller than mass of 
two nucleons (flavor-octet baryons) , $M_{\rm H} < 2M_{\rm N}$, 
in the chiral limit.

\begin{table}[h]
\centering
\caption{The mass $M_{\rm H}$ and the radius $\sqrt{\langle r^2 \rangle_{\rm H}}$ 
of the H-dibaryon in the chiral limit in holographic QCD, together with 
those of the $B=1$ hedgehog (HH) baryon.
}
\label{tab-1}      
\begin{tabular}{llll}
\hline
$M_{\rm H}$ & $\sqrt{\langle r^2 \rangle_{\rm H}}$ & 
$M_{B=1}^{\rm HH}$ & $\sqrt{\langle r^2 \rangle_{B=1}^{\rm HH}}$ \\\hline
1673 MeV & 0.413 fm & 836.7 MeV & 0.362 fm  
\\\hline
\end{tabular}
\end{table}

Finally, we examine the vector-meson effect for the H-dibaryon 
by comparing with the $\rho(x)=0$ case. 
As the result, we find that the chiral profiles $F(r)$ and $\varphi(r)$ 
are almost unchanged and slightly shrink by the vector-meson effect, 
and the energy density also shrinks slightly, as shown in Fig.~\ref{fig-4}.
As a significant vector-meson effect, 
we find that about 100MeV mass reduction is 
caused by the interaction between NG bosons and vector mesons 
in the interior region of the H-dibaryon.

\section{Summary and Concluding Remarks}
We have studied the H-dibaryon (uuddss) as the 
$B=2$ SO(3)-type topological chiral soliton solution in holographic QCD for the first time.
The H-dibaryon mass is twice of the $B=1$ hedgehog-baryon mass,
$M_{\rm H} \simeq 2.00 M_{B=1}^{\rm HH}$,
and is estimated about 1.7GeV, 
which is smaller than mass of two nucleons (flavor-octet baryons), in the chiral limit. 
In holographic QCD, we have found that 
the vector-meson effect gives a  slight shrinkage of the chiral profiles and 
the energy density, and also gives about 100MeV mass reduction of the H-dibaryon.
\section*{Acknowledgements}
The authors thank S. Sugimoto and T. Hyodo for the useful discussions with them.

\end{document}